\documentstyle[12pt,psfig,aaspp4]{article}

\newcommand{\msun}{\mbox{${\rm M}_\odot$}}
\newcommand{\msunyr}{ \mbox{ ${\rm M}_\odot \,{\rm yr}^{-1}$ } }

\newcommand{\rsun}{\mbox{${\rm R}_\odot$}}
\newcommand{\kms}{\mbox{${\rm km\,s}^{-1}$}}

\newcommand{\Wo}{\mbox{${\rm W_0}$}}
\newcommand{\zsun}{\mbox{$Z_{\odot}$}}

\newcommand{\Mwind}{\mbox{${\dot{m}_{\rm wind}}$}}
\newcommand{\Mwindsq}{\mbox{${\dot{m}^2_{\rm wind}}$}}

\newcommand{\Teff}{\mbox{${T_{\rm eff}}$}}

\newcommand{\vesc}{\mbox{${v_{\rm ce}}$}}

\newcommand{\vwind}{\mbox{$v_{\rm wind}$}}
\newcommand{\vwindsq}{\mbox{$v^2_{\rm wind}$}}
\newcommand{\Lx}{\mbox{$L_{\rm X}$}}
\def\apgt{\ {\raise-.5ex\hbox{$\buildrel>\over\sim$}}\ }
\def\aplt{\ {\raise-.5ex\hbox{$\buildrel<\over\sim$}}\ }
\def\url#1{{\tt #1}}

\lefthead{S.\ F.\ Portegies Zwart et al.}
\righthead{Chandra observations of R\,136}

\begin{document}


\title{A dozen colliding wind X-ray binaries 
       in the star cluster R\,136 in the 30\,Doradus region}

\medskip 

\author{Simon F.\ Portegies Zwart$^{1, 2, 3}$,
	David Pooley$^3$, 
	Walter, H.\ G.\ Lewin$^3$
       }

\vfill\noindent
$^1$Astronomical Institute ``Anton Pannekoek'', Univeristy of
Amsterdam, Kruislaan 403, 1098 SJ Amsterdam, NL \\
$^2$Section Computational Science, Univeristy of
Amsterdam, Kruislaan 403, 1098 SJ Amsterdam, NL \\
$^3$Massachusetts Institute of Technology, Cambridge, MA 02139, USA \\
\bigskip


Subject headings: stars: early-type --- stars: Wolf-Rayet ---
galaxies:) Magellanic Clouds --- X-rays: stars --- X-rays: binaries
--- globular clusters: individual (R136)

\newpage

\begin{abstract}

We analyzed archival {\em Chandra} X-ray observations of the central
portion of the 30 Doradus region in the Large Magellanic Cloud. The
image contains 20 X-ray point sources with luminosities between $5
\times 10^{32}$ and $2 \times 10^{35}$ erg\,s$^{-1}$ (0.2 -- 3.5\
keV).  A dozen sources have bright WN Wolf-Rayet or spectral type O
stars as optical counterparts.  Nine of these are within $\sim
3.4$\,pc of R\,136, the central star cluster of NGC\,2070.
We derive an empirical relation between the X-ray luminosity and the
parameters for the stellar wind of the optical counterpart. The
relation gives good agreement for known colliding wind binaries in the
Milky Way Galaxy and for the identified X-ray sources in NGC\,2070.
We conclude that probably all identified X-ray sources in NGC\,2070
are colliding wind binaries and that they are not associated with
compact objects.  This conclusion contradicts Wang (1995) who argued,
using {\em ROSAT} data, that two earlier discovered X-ray sources are
accreting black-hole binaries.
Five of the eighteen brightest stars in R\,136 are not visible in our
X-ray observations. These stars are either single, have low mass
companions or very wide orbits.  The resulting binary fraction among
early type stars is then unusually high (at least 70\%).  

\end{abstract}

\section{Introduction}
30 Doradus (NGC\,2070) is an active star forming region in the Large
Magellanic Cloud (LMC). Its most striking feature in the optical is the
young and compact star cluster R\,136 (HD\,38268; see Walborn
1973).\nocite{1973ApJ...182L..21W} The cluster was long thought to be
a single star with a mass exceeding 3000\,\msun\ (Cassinelli et al.\,
1981).\nocite{1981Sci...212.1497C} The discovery that this ``single
star'' fitted a King (1966) model indicated that the object is in fact
a cluster of stars (Weigelt \& Baier
1985).\nocite{1985A&A...150L..18W} Later the {\em Hubble Space
Telescope} provided direct proof for this hypothesis by resolving the
cluster into individual stars (Campbell et al.\, 1992; Massey \& Hunter
1998).\nocite{1992AJ....104.1721C}\nocite{1998ApJ...493..180M}

The structural parameters of R\,136 --- mass (21\,000 --
79\,000\,\msun), half-mass radius ($\sim 0.5$\,pc), and density
profile ($\Wo\sim 6$, see e.g. Campbell et al.\,
1992;\nocite{1992AJ....104.1721C} Brandl et al.\,
1996;\nocite{1996ApJ...466..254B} and Massey \& Hunter
1998)\nocite{1998ApJ...493..180M} --- are quite similar to three well
known counterparts in the Milky Way Galaxy: the Arches cluster (Object
17, Nagata et al.\ 1995)\nocite{1995AJ....109.1676N}, the Quintuplet
(AFGL\,2004, Nagata et al.\ 1990; Okuda et al.\,
1990)\nocite{1990ApJ...351...83N}\nocite{1990ApJ...351...89O}, and
NGC\,3603 (Brandl 1999).\nocite{1999A&A...352L..69B} All being younger
than $\sim 3$\,Myr, these clusters are also comparable in age.
NGC\,3603 is hidden behind a spiral arm, but is much easier to observe
than the Arches and the Quintuplet, which are located near Galactic
coordinates $l=0.122^\circ$, $b=0.018^\circ$ ($\sim 35$\,pc in
projection) of the Galactic center.  The extinction in the direction
of the Galactic center easily exceeds 20 magnitudes in visual.  R\,136
is conveniently located in the LMC and therefore is quite unique as it
is neither obscured by dust and gas nor perturbed by external tidal
fields; the tidal effect of the LMC is negligible (see Portegies Zwart
et al.\, 1999). The large distance of about 50 kpc, however, limits
the observability of the cluster. Note that the star clusters
Westerlund 1 and 2 (Westerlund 1960, 1961) both have rather similar
characteristics as the above mentioned clusters.

In 1991, Wang \& Helfand\nocite{1991ApJ...370..541W} studied the 30
Doradus region with the {\em Einstein} Imaging Proportional
Counter. They found that hot gas surrounds a $\sim 300$\,pc area
around R\,136. They also found two marginally significant point
sources with the {\em Einstein} High Resolution Imager in the central
area of 30 Doradus. A follow-up with the {\em ROSAT} High Resolution
Imager confirmed the presence of
two point sources.  Wang (1995) analyzed these two sources and used
their similarities with Cyg X-1, LMC X-1 and LMC X-3 to conclude that
the observed X-ray sources in 30 Doradus also host black holes.
Recent XMM observations by Dennerl et al.\, (2001) showed a wealth of
X-ray sources in the 30 Doradus region. They pay little
attention, however, to the X-ray point sources in R\,136.  In a preliminary
study of {\em Chandra} data of R\,136, Feigelson (2000, see also Townsley et
al 2001, in preparation) mentions the presence of a dozen bright X-ray
sources with O3 or WN star companions. He argues that the dimmer
sources could be colliding wind binaries, but that the brighter
sources are probably X-ray binaries with neutron star or black
hole primaries.  From a practical point of view, however, there are some
difficulties associated with claims that there are neutron stars or
black holes in R\,136.

The most massive stars in R\,136 exceed 120\,\msun. Massey \& Hunter
(1998) took high resolution spectra of the brightest 65 stars and
concluded that their age is $<1$--2\,Myr, where the intermediate mass
stars are slightly older 4--5\,Myr (Sirianni et al.\,
2000).\nocite{2000ApJ...533..203S} They also argue that the high
abundance of stars with spectral type O3 and earlier is a consequence
of the young age of the cluster. de Koter et al.\,
(1997)\nocite{1997ApJ...477..792D} analyzed the spectra of the most
massive stars in R\,136 and concluded that the ages of these stars are
$\aplt 1.5$\,Myr, which is consistent with the results by Massey \&
Hunter (1998).

The hydrogen and helium burning age of a 120\,\msun\ star depends
quite sensitively on its metalicity $Z$, being as low as 2.87\,Myr for
a $Z=0.001$ (the solar metalicity $\zsun = 0.02$) to 4.96\,Myr for
$Z=0.04$ (Meynet et al.\, 1994).\nocite{1994A&AS..103...97M} Because
R136 is only $\sim 2$\,Myr old, no star in it can have left the main
sequence yet.  It would therefore be quite remarkable if the star
cluster were able to produce black holes before 2\,Myr.

We analyze archival {\em Chandra} data of the central portion of the
30 Doradus region and confirm the detection of the two X-ray point
sources previously found by Wang \& Helfand (1991). In addition, we
find 18 new X-ray sources, of which 13 have WN Wolf-Rayet stars or
early type O3f$^\star$ stars as optical counterparts.  One of these
X-ray sources is probably a blend of several WN stars in the
sub-cluster R\,140 (to the north of R\,136).  We argue that the X-rays
produced by these systems originate from the colliding winds of the
early type stars. The X-ray luminosities of three of these stars are
an order of magnitude higher than the X-ray luminosities of colliding
wind systems in the Milky Way Galaxy.

In Sect.\,\ref{Sect:Obs} we describe the {\em Chandra} data and
analysis, followed by a description of our findings in
Sect.\,\ref{Sect:Results}. We discuss our results in
Sect.\,\ref{Sect:Disc}.

\section{Observations and data analysis}\label{Sect:Obs}

The 30~Doradus region was observed with {\it Chandra} for 21~ksec on
2000~Sep~13 with the Advanced CCD Imaging Spectrometer-Imager
(ACIS-I), a $16\farcm9\times 16\farcm9$ array of four front-side
illuminated CCDs (Garmire, Nousek, \& Bautz in preparation). The
R\,136 star cluster was at the telescope's aim point.  The data were
taken in timed-exposure (TE) mode using the standard integration time
of 3.2~sec per frame and telemetered to the ground in very faint (VF)
mode.  See the {\it Chandra} Proposers' Observatory Guide for details
({\it Chandra} website \url{http://asc.harvard.edu/}).

We followed the data preparation threads provided by the {\it Chandra}
team and given on their website.  We used the {\it Chandra}
Interactive Analysis of Observations (CIAO) software package to
perform the reductions, with the most up-to-date (as of June 2001)
calibration files (gain maps, quantum efficiency, quantum efficiency
uniformity, effective area) available for this observation.  Bad
pixels were excluded.  Intervals of bad aspect and intervals of
background flaring\footnote{See
\url{http://asc.harvard.edu/cal/Links/Acis/acis/Cal\_prods/bkgrnd/current}
for a discussion of background flares.} were searched for, but none
were found.

\begin{figure}[htbp!]
\hspace*{1cm}
a) \psfig{figure=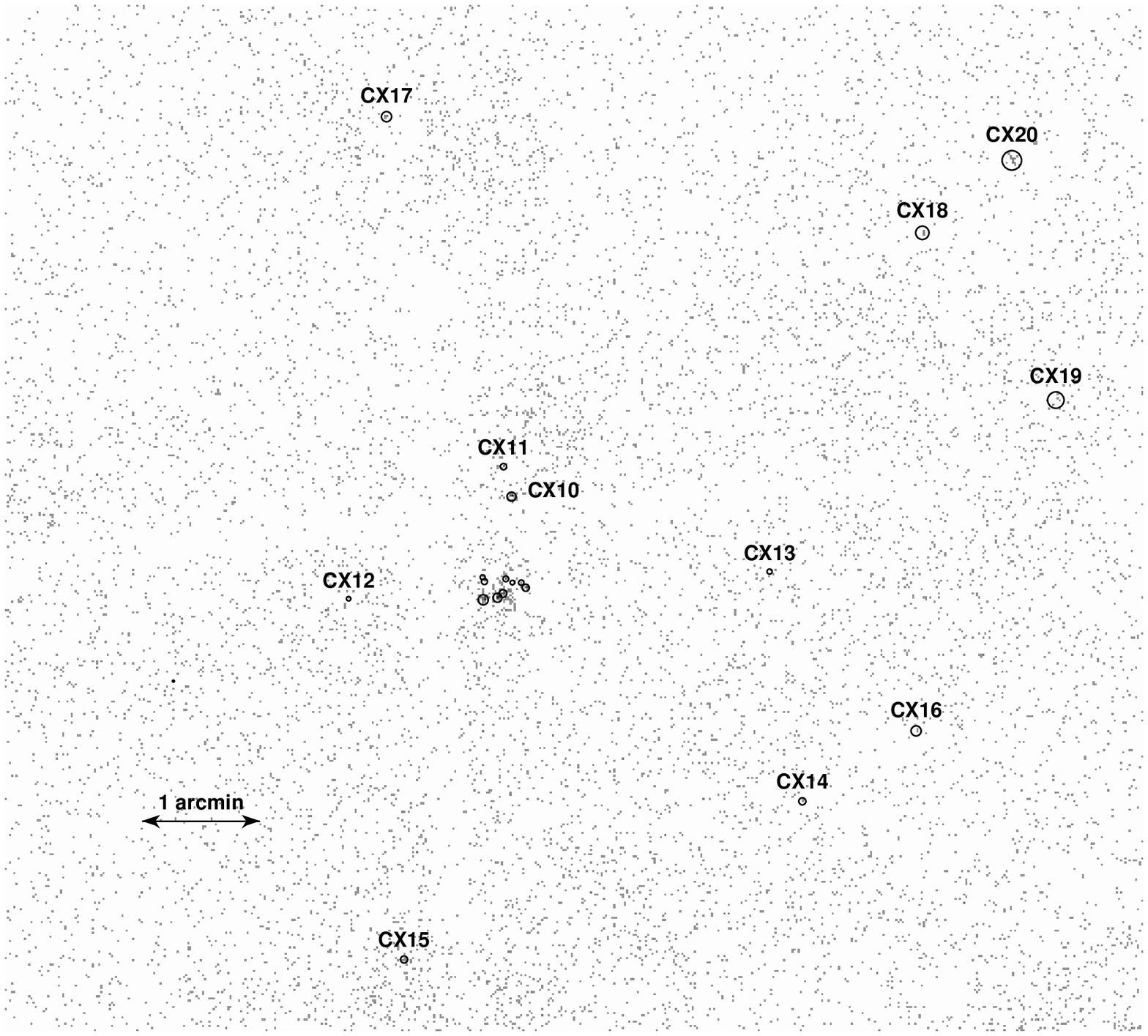,width=7cm,angle=0}
b) \psfig{figure=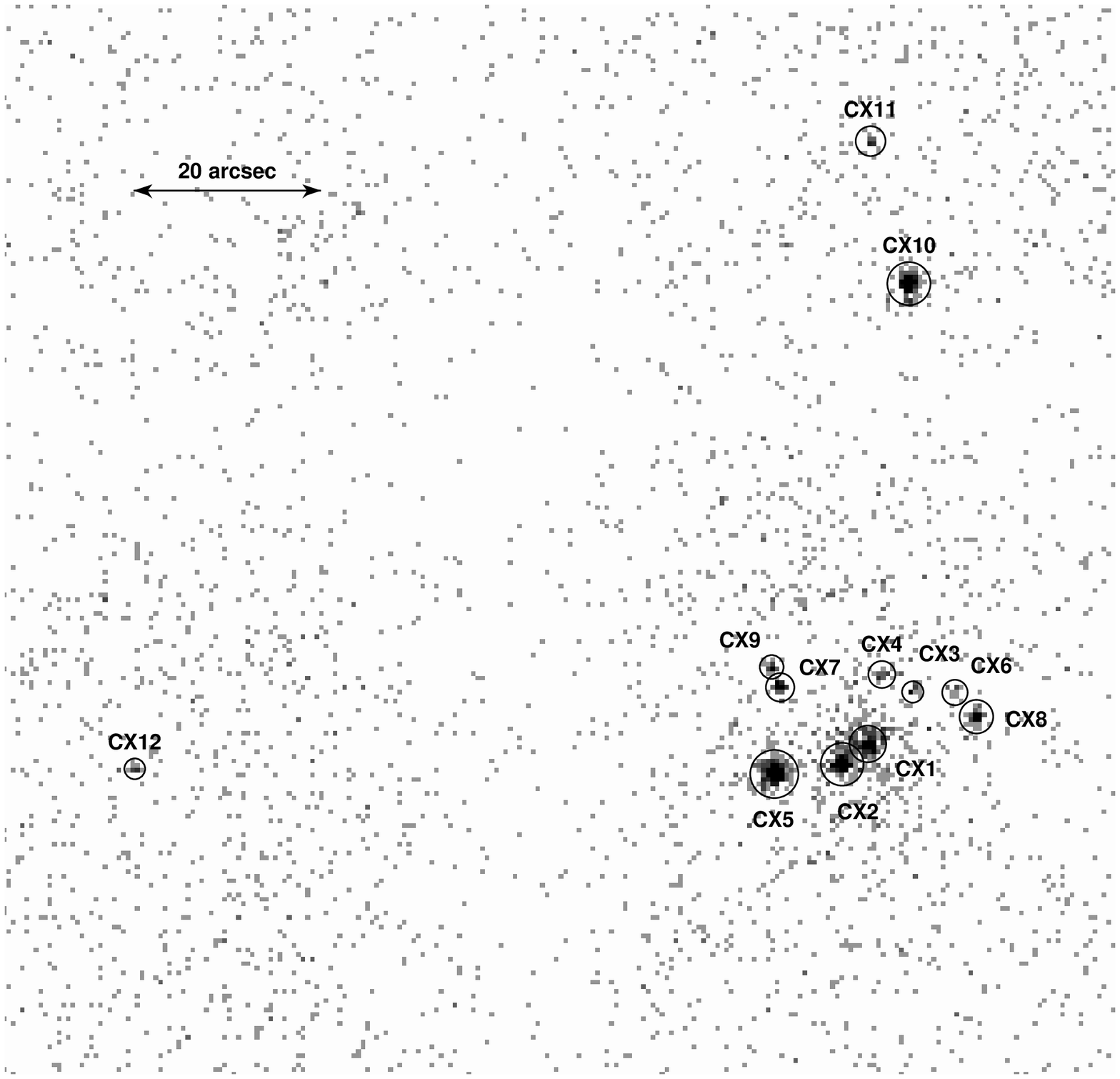,width=7cm,angle=0}
\caption[]{X-ray image of the star cluster
R\,136 in the 30 Doradus region. 
a) $11' \times 11'$ view, b) close up of the cluster.
The sources are identified in Tab\,\ref{Tab:Obs}.}
\label{Fig:Xray}
\end{figure}

To detect sources, we first filtered the unprocessed data to exclude
software-flagged cosmic ray events and then processed the data without
including the pixel randomization that is added during standard
processing.  This custom processing slightly improves the point spread
function (PSF).  We then used the CIAO tool {\it wavdetect}, a wavelet
based source detection program.  We found 30 point sources and two
extended sources (SNR 0538-69.1 and an unidentified one) in the entire
ACIS-I field.  Fig.\,\ref{Fig:Xray} shows a $11' \times 11'$ image
(left) and a blow up of the central 7$''$ square (right) {\em Chandra}
image with the source names identified. The names of the sources
correspond to the 20 sources listed in Tab.~\ref{Tab:Obs}.

Radial surface brightness profiles were constructed for the four
brightest sources and compared to expected PSF profiles for those chip
locations.  Only CX10 (near the star cluster R\,140) was inconsistent
with a point source (3.4~$\sigma$).

\begin{figure}[htbp!]
\hspace*{1cm}
\psfig{figure=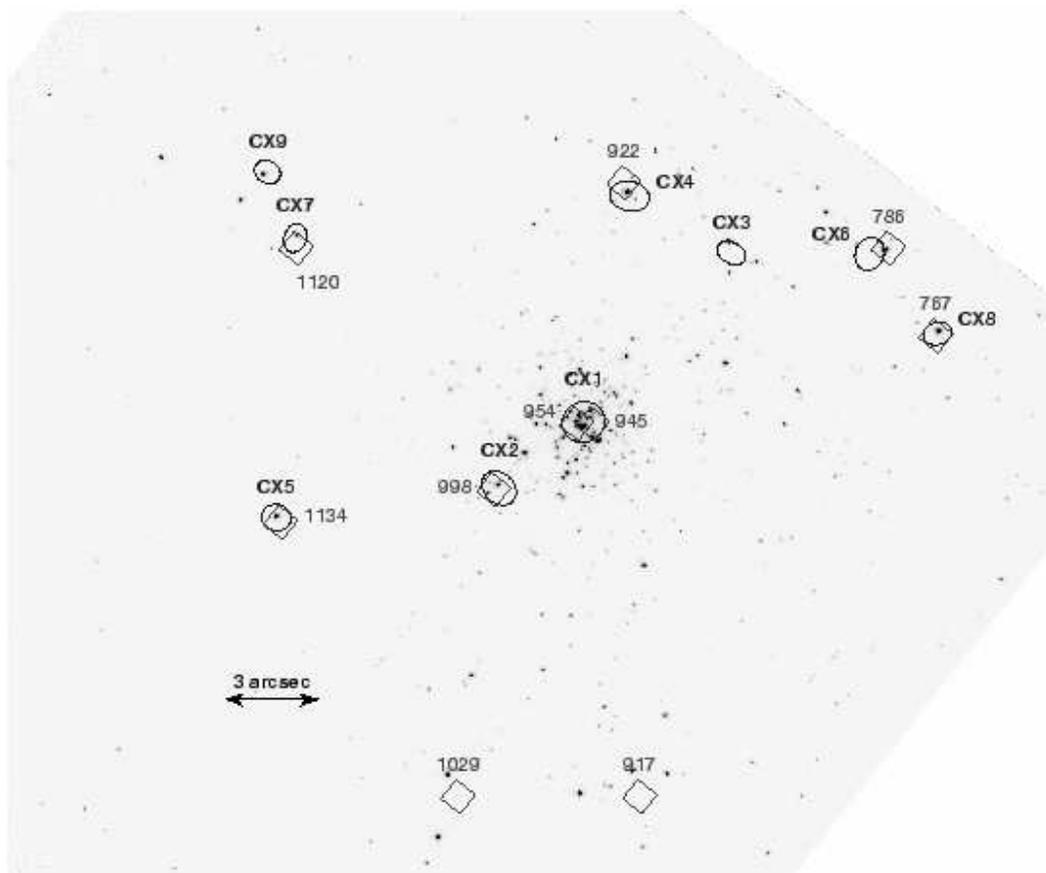,width=14cm,angle=0}
\caption[]{Hubble Space Telescope image of R\,136, the central portion
of the 30\,Doradus region (Massey \& Hunter
1998).\nocite{1998ApJ...493..180M} The exposure time was 26 seconds
with filter F336W (centered on 3344\,\AA, and with a 381\,\AA\
bandpass) using WFPC2.  The Wolf-Rayet stars identified by Parker
(1993) are indicated by squares and the ellipses give the locations of
the Chandra X-ray sources (see Tab.\,\ref{Tab:Obs} and
\,\ref{Tab:OC}).  The sizes of the ellipses give the $1\sigma$
positional accuracy for the {\em Chandra} sources.  }
\label{Fig:Obs}
\end{figure}

The positions listed in Tab.~\ref{Tab:Obs} are the {\it wavdetect}
centroid positions based on the {\it Chandra} astrometric solution.
The uncertainties range from $\sim$0\farcs1 for the brighter sources
to $\sim 1^{\prime\prime}$\ for the dimmer sources.  To perform source
identifications, we used Parker's catalog of bright stars in the
30~Doradus region (Parker 1993).  These positions are accurate within
0\farcs4.  Registration of the {\it Chandra} frame of reference with
the Parker frame was accomplished via a uniform shift calculated from
a least-squares best fit to the offsets from nearby optical sources of
eight X-ray sources (CX2, CX4, CX5, CX6, CX7, CX8, CX12 and CX19), six
in the central region and two outside the central region.  In a
similar manner, the Parker frame was registered with the {\it Hubble}
frame by a uniform shift based on five stars (names from Parker 1993:
1120, 1134, 922, 786, 767) in the central region.  The Parker-to-{\it
Hubble} shift was 0\farcs 07 in RA (true arcsec) and $-$1\farcs32 in
Dec.  The total {\it Chandra}-to-{\it Hubble} shift was
1\farcs35 in RA (true arcsec) and 0\farcs 66 in Dec.
Fig.~\ref{Fig:Obs} shows the {\it Chandra} sources (ellipses) and
Parker stars (boxes) overlaid on a {\it Hubble} image of the central
region.  The sizes of the ellipses and boxes indicate the 1~$\sigma$
position errors.

\begin{table*}[htbp!]
\caption[]{ {\em Chandra} detections of X-ray point sources near the
star cluster R\,136. RA: $5^h38^m43^s.2$, Dec: $69^\circ 6'0''$ in the
year 2000AD.  Columns give the name of the source, its position in RA
and Dec, the distance to the star R\,136a (CX1), number of counts in
0.3--8.0~keV, the fit parameters $kT_{\rm mekal}$ and $n_H$, the
Chi-squared $\chi^2$ statistic and degrees of freedom (dof), and the
unabsorbed luminosity in the 0.2 to 3.5\,keV {\em Einstein} band.  For
sources with $\aplt 90$ counts (including CX20 because of the lack of
an optical counterpart) we estimate the luminosities based on a
best-fit linear relation between counts and luminosities for four of
the five brightest sources (we excluded CX10, see
sect.\,\ref{Sect:Obs}).}
\begin{flushleft}
\begin{tabular}{l|rlr|rllllll}
\\ \hline
Source & RA&Dec& $r$ &Counts & $kT_{\rm mekal}$ & $n_H$ &
	$\chi^2/$dof & $\log L_{\rm 0.2-3.5 keV}$ \\ 
& & &[$''$]& &[keV] & [$10^{21}$ cm$^{-2}$] & & [erg s$^{-1}$] \\ \hline

CX1 & 05:38:42.210& -69:06:03.85&0   &  146& 1.7& 2.8  & 14.6/13 & 34.34 \\ 
CX2 & 05:38:42.722& -69:06:06.02&3.5 &  390& 2.1& 6.3  & 19.7/18 & 34.93 \\ 
CX3 & 05:38:41.311& -69:05:58.30&7.3 &   13&    &      &         & 33.17 \\ 
CX4 & 05:38:41.930& -69:05:56.43&7.6 &   16&    &      &         & 33.31 \\ 
CX5 & 05:38:44.077& -69:06:07.05&10.5&  997& 3.3& 4.8  & 33.0/32 & 35.26 \\ 
CX6 & 05:38:40.470& -69:05:58.34&10.8&    9&    &      &         & 32.86 \\ 
CX7 & 05:38:43.962& -69:05:57.81&11.2&   51&    &      &         & 33.93 \\ 
CX8 & 05:38:40.042& -69:06:00.94&12.0&   91& 2.1& 1.6  & 8.0/6   & 34.00 \\ 
CX9 & 05:38:44.130& -69:05:55.64&13.2&   20&    &      &         & 33.44 \\ 
CX10& 05:38:41.391& -69:05:14.67&49.4&  374& 1.1& 7.9  & 11.4/17 & 35.22 \\ 
CX11& 05:38:42.156& -69:04:59.46&64.4&   17&    &      &         & 33.35 \\ 
CX12& 05:38:56.852& -69:06:06.48&78.4&   11&    &      &         & 33.04 \\ 
CX13& 05:38:16.909& -69:05:52.57&136 &    6&    &      &         & 32.23 \\ 		    
CX14& 05:38:13.755& -69:07:49.13&185 &    7&    &      &         & 32.55 \\ 
CX15& 05:38:51.602& -69:09:09.59&192 &   15&    &      &         & 33.27 \\	   	 
CX16& 05:38:02.963& -69:07:13.28&221 &    8&    &      &         & 32.73 \\	   	 
CX17& 05:38:53.226& -69:02:01.82&249 &   13&    &      &         & 33.17 \\ 
CX18& 05:38:02.508& -69:03:00.44&281 &   22&    &      &         & 33.50 \\	   	 
CX19& 05:37:49.811& -69:04:25.14&297 &   17&    &      &         & 33.35 \\ 
CX20& 05:37:54.057& -69:02:23.54&339 &   94&    &      &         & 34.22 \\
\hline
\end{tabular}
\end{flushleft}
\label{Tab:Obs} 
\end{table*}

\begin{figure}[htbp!]
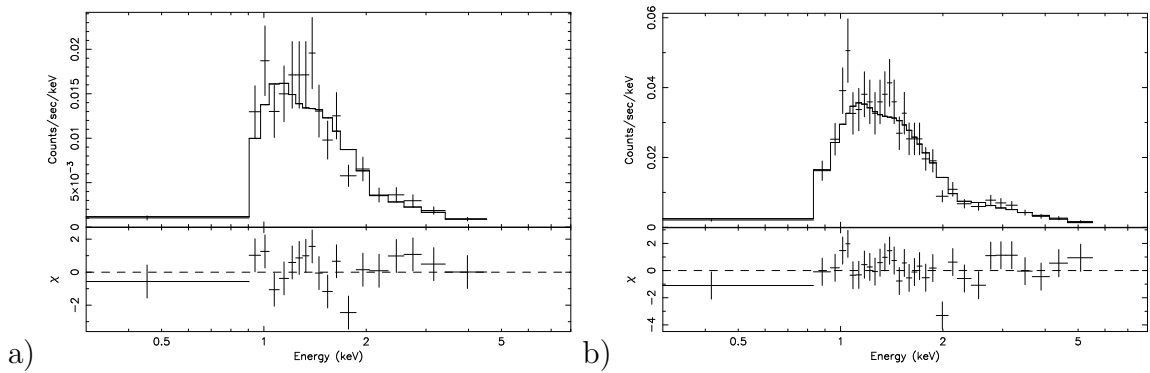

\hspace*{1cm} a) \psfig{figure=F3a.ps,width=7cm,angle=-90} b)
\psfig{figure=F3b.ps,width=7cm,angle=-90}
\caption[]{X-ray spectra of the two brightest sources CX2 (a) and CX5
(b). The data (crosses) and best-fit {\em mekal} model (solid line)
are shown in both panels.  The lower panels give the contribution to
the $\chi^2$ statistics for each bin.  }
\label{Fig:spectra}
\end{figure}

X-ray spectra of the 5 brightest {\it Chandra} sources were extracted
with the CIAO tool {\it dmextract}.  Spectral fitting (using $\chi^2$
statistics) was done in XSPEC with the data grouped to $\geq$10 counts
per bin for CX1 and CX8, $\geq$20 counts per bin for CX2 and CX10, and
$\geq$30 counts per bin for CX5.  Because of the association of nearly
all the X-ray sources with Wolf-Rayet stars, we chose the {\it mekal}
model (Mewe et al.\, 1985) with absorption, which fits better than
thermal bremsstrahlung models.  A summary of the results is given in
Tab.\ref{Tab:Obs}.  The spectra of the brightest two sources (CX2 and
CX5), which are quite representative, are shown in
Fig.~\ref{Fig:spectra}.  For a better comparison with previous
observations, which were mostly done with the {\em Einstein}
observatory, we calculate the luminosities over the same band.  Though
a source can be detected with only a few counts, $\apgt 100$ counts
are required to fit a spectrum with any confidence.  To estimate the
luminosity of the dimmer sources, we use a best-fit linear
relationship between observed counts and unabsorbed luminosities for
four of the brightest five sources.  Because the source CX10 is bright
for its count rate and extended, we excluded it when deriving this
relation.

We identify source CX10 with W5 and CX5 with W7 from Wang (1995).  The
luminosities for these two sources calculated from our best fit models
are lower than those of Wang by a factor of $\sim$10, but this is
somewhat deceptive.  Wang used multicolor blackbody disk models (MBD)
with absorption and could only fit over the 0.5--2~keV {\it ROSAT}
band.  Such models do not fit our broader band (0.3--8~keV) data well.
If we restrict our data to the {\it ROSAT} band, we find that MBD
models fit the data, but the column densities required for good fits
with these models are much higher than those required for fits with
the {\it mekal} model.  When we calculate {\it intrinsic} source
luminosities based on MBD fits to the 0.5--2~keV {\it Chandra} data,
our results are consistent with those of Wang. There is no evidence
for variability on the time scale between the two observations 9 years
apart.  Also, none of our sources show statistically significant
variability during the {\it Chandra} observation.

\section{Results and interpretation}\label{Sect:Results}

\subsection{Optical counterparts}

We concentrate on the 11 X-ray sources (CX1--CX11) which are located
within 100$''$ ($\sim 25$\,pc in projection at a distance of 51\,kpc)
from the central star R\,136a1 of the star cluster R\,136.  For the
other eight sources we have insufficient information to include them
in our study. Two of these remaining sources (CX12 and CX17) have a WN
type Wolf-Rayet star as optical counterparts (see Tab.\,\ref{Tab:OC}),
CX14 is an IRAS source and CX19 has the foreground star MH\,2 as
counter part (Fehrenback and Duflot, 1970), which is observed in
infrared (McGregor \& Hyland, 1981) and classified as a spectral type
M star (Hyland, Thomas \& Robinson 1978).  The remaining five sources
(CX13, CX15, CX16, CX18 and CX20) are not identified.

All 11 sources (CX1 to CX11) have a bright optical star within the
$1\sigma$\ error ellipse of the corrected {\em Chandra} position.
Fig.\,\ref{Fig:Obs} gives a {\em Hubble Space Telescope} image of
R\,136 (Massey \& Hunter 1998).  The ellipses indicate the $1\sigma$
error at the location of our 11 {\it Chandra} point sources near
R\,136.  Details about these sources are given in Tab.\,\ref{Tab:Obs}.

Parker (1993)\nocite{1993AJ....106..560P} identifies 12 WN type
Wolf-Rayet stars (see Tab.\,\ref{Tab:OC}) near the star cluster
R\,136. These are identified in Fig.\,\ref{Fig:Obs} by squares.  Five
of these (R\,136a1, a2, a3, a5 and R\,136b) are embedded deep in the
core of the cluster. The source CX10 has the small cluster R\,140 as
counterpart, which is at about 11.5\,pc in projection to the north of
R\,136a. This cluster contains at least two WN stars and one WC star
(Moffat 1987). We found CX10 to be extended and it may well be a blend
of several colliding wind binaries.

Seven of nine of the remaining sources have WN stars as counterparts,
and the other two coincide with spectral type O3f$^\star$ stars
(Crowther \& Dessart, 1998).\nocite{1998MNRAS.296..622C} Melnick (1978
see also Melnick 1985 and Bosch et al.\ 1999) lists most of the
Wolf-Rayet stars (and one of the O3f$^\star$ stars which are
associated with our X-ray sources) as binaries with a spectral type O
or B companion.  The four exceptions P860, Mk39, R134 and R139 are not
known to be binaries. The star Mk33Na is listed as a binary by
Crowther \& Dessart (1998) but not by Melnick (1978; 1985).  The two
stars R\,144 and R\,145, which are associated with CX17 and CX12,
respectively, are little discussed in the literature, both are WN6h
stars (Crowther \& Smith\ 1997).\nocite{1997A&A...320..500C}

Tab.\,\ref{Tab:OC} lists the X-ray sources and their counterparts.

\begin{table*}[htbp!]
\caption[]{
%
Optical counterparts for the X-ray sources.  The stellar numbering P
(column 3) is from Parker (1993).  The spectral type and mass loss
rates (columns 4 and 5) are taken from Crowther \& Dessart (1998, and
private communication from Paul Crowther). Bolometric luminosities,
masses and escape velocities ($v_{\rm esc}$) for the O stars are from
Massey \& Hunter using the calculations of Vacca et al.\
(1996).\nocite{1996ApJ...460..914V} Column 8 ($v_{\rm esc}$) gives,
for the Wolf-Rayet stars, the escape velocity from the stellar core
(\vesc) as described in Sect.\ 3.3.  The terminal wind velocities for
the O stars are calculated from the effective temperature, as given by
Crowther \& Dessart. The wind velocities of the Wolf-Rayet stars are
calculated assuming the following compositions, for WN5/6 stars we
adopt $Y=0.983$ and $Z=0.0172$, for stars with hydrogen enhancement we
use $Y=0.7324$ and $Z=0.0176$.  The last column contains some notes
about the binarity or alternative classification from Melnick (1978;
1985)\nocite{1978A&AS...34..383M}\nocite{1985A&A...153..235M} and from
Bosch et al.\ (1999)\nocite{1999A&AS..137...21B}.  Uncertain values
are indicated by a question mark (?) and unknown values are left
blank.  }
\begin{flushleft}
\begin{tabular}{l|rrl|clcrrc}
\\ \hline
Source& star & P    & Spectral   &$\log \Mwind$& $\log L_{\rm bol}$ 
                    & mass & $v_{\rm esc}$ & $v_\infty$& note \\
      &      &      & type [CD]  & [\msunyr]& [erg\,s$^{-1}$]  
		    & [\msun]&\multicolumn{2}{c}{[1000 \kms]} & \\
\hline	     	   
CX1 &  R136a &      & 3 WN5      & -4.4?&40.2&111 &1.93&1.08& WN4.5+OB \\
CX2 &  R136c & 998  & WN5h (+?)  & -4.4 & 40.0 &128  &2.67&1.45& WN7+OB \\
CX3 &  ...   & 860  & O3V        & -5.5 & 30.5 & 77  &3.70&    & O7Vf \\
CX4 &  Mk42  & 922  & WN6/O3If   & -4.3&40.0 & 66.5&1.57&0.93& WN+OB \\
CX5 &  Mk34  & 1134 & WN5h       & -4.3 & 40.1 &179  &3.14&1.66& WN4.5 (+O?)\\
CX6 &  R134  & 786  & WN6(h)     & -4.1 & 40.1 &179  &3.14&1.66& WN7 \\
CX7&Mk33Sa& 1120 & O3IIIf$^\star$+WC5& -5.3 &39.5 & 105&2.34&  & WC5+O4 \\
CX8 &  Mk39  & 767  & WN6/O3If   & -4.4 &40.0&66.5 &1.57&0.93& O4If \\
CX9 &  Mk33Na& 1140 &O3If$^\star$+O6.5V&-5.1 &39.8 &105 &2.37&   & WN4/O4 \\
CX10&  R140a & 880  & WN6        & -4.4?& 39.9 & 53.7&1.48&0.91& WC5+WN4 \\
CX11&  R139  & 952  & O6Iaf/WN   & -4.1?& 40.0 & ?   &2.40&    & \\
CX12&  R145  &      & WN6h       & -3.84& ?    & \\
CX14&  IRS 2 &      & \\		         	 
CX17&  R144  &      & WN6h       & -3.65& ?    & \\ 
CX19&  MH\,2 &      &            & \\
\hline 	     	  
\hline
\end{tabular}
\end{flushleft}
\label{Tab:OC} 
\end{table*}

\subsection{The nature of the X-ray sources}

\begin{figure}[htbp!]
\hspace*{1cm}
\psfig{figure=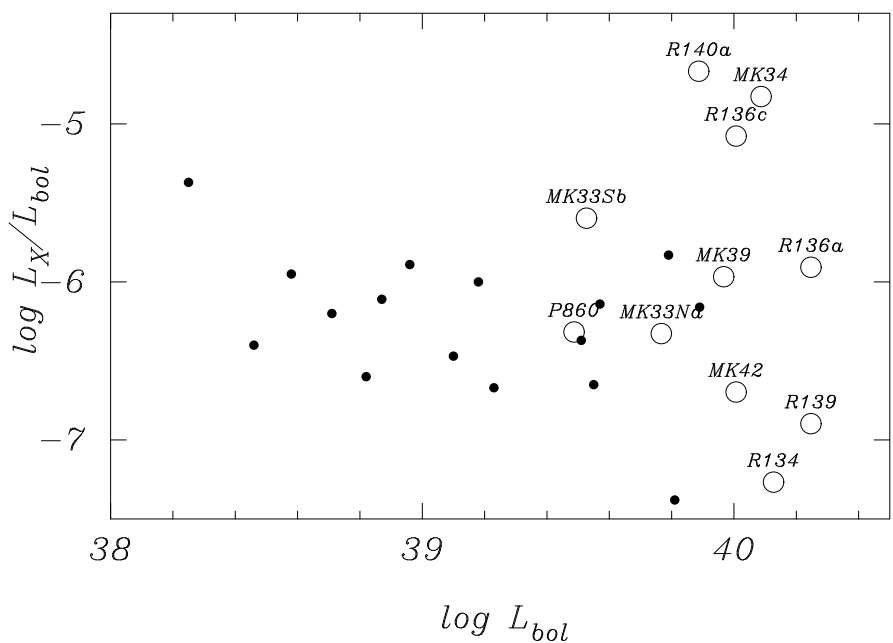,width=14cm,angle=-90}
\caption[]{Ratio of the X-ray luminosity (0.2 -- 3.5 keV) to the
Bolometric luminosity ($\Lx/L_{\rm bol}$ for the X-ray sources from
Tab.\,\ref{Tab:OC} and Tab.\,\ref{Tab:KnownWR} as a function of the
Bolometric luminosity (in erg\,s$^{-1}$).  The known Galactic
colliding wind binaries from Tab.\,\ref{Tab:KnownWR} are indicated
with bullets (see Fig.\,\ref{Fig:fLx_WR} for the error bars).  The
sources from Tab.\,\ref{Tab:OC} are indicated by circles and
identified by the names of their optical counterparts.  Note that the
two types of sources (Galactic versus those near NGC\,2070) are
selected on different criteria and a direct comparison has to be
carried out with care.  }
\label{Fig:LxLb_ratio}
\end{figure}

We now study the possibility that the X-rays are produced in the
stellar winds in the optical counterparts.  Wolf-Rayet stars are known
to be bright X-ray sources (Pallavicini et al.\,
1981),\nocite{1981ApJ...248..279P} which follow the empirical relation
between X-ray and Bolometric luminosity of $\Lx/L_{\rm bol} \simeq
10^{-7\pm 1}$.  This relation holds for a wide range of stellar
spectral types down to late B.  White \& Long
(1986)\nocite{1986ApJ...310..832W} confirm the existence of this
relation for WN and WC type Wolf-Rayet stars.  An extensive study by
Pollock (1987)\nocite{1987ApJ...320..283P} using the {\em Einstein}
observatory confirms this relation for binaries with a WN or WC
Wolf-Rayet star as primary.

The sample of Chlebowski \& Garmany (1991) contains early type O
stars, which may follow a different relation between $\Lx$ and $L_{\rm
bol}$ than Wolf-Rayet stars. However, the Wolf-Rayet stars listed by
White \& Long (1986, 2 WN and 4 WC stars) and by Pollock (1987, 7 WN
and 4 WC stars) follow the same relation as the early type O stars of
Chlebowski \& Garmany (1991). (To minimize confusion we decided to
show only the stars listed by Chlebowski \& Garmany in
Fig.\,\ref{Fig:LxLb_ratio}.)

In Fig.\,\ref{Fig:LxLb_ratio} we compare the relation between
Bolometric luminosity $L_{\rm bol}$ and $L_X/L_{\rm bol}$ of 11 of our
observed X-ray sources with that of the 16 X-ray bright stars in the
sample discussed by Chlebowski \& Garmany (1991).  To make the
comparison, we calculate the 0.5--3.5\,keV luminosity from our
best-fit models (which were fit over the entire 0.3--8.0\,keV {\em
Chandra} band) since this was the band used by Chlebowski \& Garmany
(1991), who observed with the {\em Einstein} observatory (see
Tab.\,\ref{Tab:Obs}).  Between the stars in NGC\,2070 and the sample
of Chlebowski \& Garmany (1991), three stars (R140a, Mk34 and R136c)
are considerably brighter than expected from the empirical relation,
and the scatter among the stars in NGC\,2070 is larger than for the
sample of Galactic sources.  The unusual brightness of CX10 may be
explained as a blend of several objects in the sub cluster R140.

We have no ready explanation for the extreme brightness of the other
two stars, but bear in mind that the stars associated with our X-ray
sources are unusually bright and massive, ranging from 50\,\msun\ to
150\,\msun\ (see Tab.\,\ref{Tab:OC}), unlike any stars in the Milky
Way Galaxy. It is therefore possible that they do not follow the same
relation as the rather ``low'' mass (15\,\msun\ to 30\,\msun) stars in
the Milky Way Galaxy. Note also that a direct comparison between the
Galactic X-ray sources and those near NGC\,2070 has to be carried out
with care as both have been selected on different grounds.

\subsection{X-rays from colliding stellar winds}

The typical mass loss rate for a WN Wolf-Rayet star is between
$10^{-5}$ and $10^{-4}$\msunyr and somewhat smaller for an Of$^\star$
star (see Tab.\,\ref{Tab:OC}).  When the fast mass outflow of one of
these stars collides with the mass outflow of a companion star, strong
shocks form. For a single star, the shocks will be much weaker because
the relative velocity of the shocked material is much smaller for a
single star than one in a binary.  The temperature in the shocks can
be very high, at which point X-rays are produced (Prilutskii \& Usov
1976; Cherepashchuk 1976; Cooke, Fabian \& Pringle
1978).\nocite{1978Natur.273..645C} The X-ray emission observed from
such systems is generally rather soft (kT$\sim 1$\,keV), originating
from the outer parts of the wind (Pollock 1987).

We estimate the total X-ray luminosity in the colliding wind in the
limit of completely isothermal radiative shocks and assuming that the
flow is adiabatic\footnote{We refer to the total X-ray luminosity as
our model is too simple to identify any energy dependency.}.  The
total X-ray luminosity is given by the product of the emissivity per
unit volume, $n^2\Lambda$, with the volume of the shock-heated wind.
Here $\Lambda \propto T^{-0.6}$ is the emission rate at which the gas
cools (see Stevens, Blondin \& Pollock 1992).  In the adiabatic limit
for a binary star with orbital separation $a \gg r$ the volume of the
shock heated gas scales as $a^3$.  Here $r$ is the stellar radius.
The density $n \propto \Mwind/(a^2\vwind)$. The total luminosity due
to the shock is then $L \propto \Mwindsq/(\vwindsq a)$.  We assuming
that the post shock temperature equals the temperature on the line
connecting the two stars and using $T \propto \vwindsq$, and define
(in cgs units):
\begin{equation}
	S_{\rm X} = {\Mwindsq \over v^{3.2}_{\rm wind} a}.
\label{Eq:Lxexp}
\end{equation}
Here we assume momentum balance of the two winds on the connecting
lines between the two stars, i.e., the two stars are equally windy
(equipetomaniac). This relation has been tested with detailed
calculations by Luo, McCray \& Mac Low (1990) and Stevens, Blondin \&
Pollock (1992).

We will now calculate the terminal velocity of the stellar wind of
early type O stars and Wolf-Rayet stars.  The luminosity $L$ of
massive main-sequence stars $L \propto m^3$ and the radius $r \propto
m$. With $L \propto r^2T^4_{\rm eff}$ and the escape velocity from the
stellar surface $v^2_{\rm esc} \propto m/r$ we can write $v_{\rm esc}
\propto T_{\rm eff}/m^{1/4}$, i.e., the escape velocity of a massive
star is proportional to its effective temperature.  The terminal
velocity in the stellar wind $v_\infty \propto v_{\rm esc}$.  However,
the velocity of the stellar wind at the moment it collides may be
smaller than the escape velocity from the stellar surface, but larger
then the terminal velocity.  We write the velocity at the moment the
winds collide as
\begin{equation}
	\vwind = v_\infty \ \left(1 - {r \over a} \right)
\label{Eq:vwind}\end{equation}
For simplicity we assume that the winds collide at distance $a$ from
the primary star with radius $r$, which we calculate from the
effective temperature and the luminosity.


We calibrate the relation between the terminal velocity and the stellar
temperature with the list of O and B stars in Tab.\,\ref{Tab:KnownWR},
which results in
\begin{equation}
	v_\infty \simeq 130 {\Teff \over [1000K]} - 2800 [\kms], 
\label{Eq:vwindOB}\end{equation}

For the O stars in our sample we can apply the same relation between
the wind velocity and the effective temperature of the star.  For
Wolf-Rayet stars, however, this relation breaks down\footnote{The
radius of a Wolf-Rayet star is ill defined because the wind is
optically thick due to the large mass loss rate; the stellar radius
depends on the wavelength. The effective temperature of a Wolf-Rayet
star is therefore ill defined.}.

Crowther \& Dessart (1998) list mass loss rates, \Mwind\ and
Bolometric luminosities $L_{\rm bol}$ for most of the optical
counterparts to our X-ray sources but not their terminal
velocities. We derive the escape velocity \vesc\ from the core of a
Wolf-Rayet star using its luminosity.  The terminal velocity then
follows from the empirical relation of Nugis \& Lamers (2000). The
wind velocity is then calculated using Eq.\,\ref{Eq:vwind}.

The terminal velocity $v_\infty$\ is given by (Nugis \& Lamers 2000)
\begin{equation}
	\log v_\infty/\vesc \simeq 0.61 - 0.13 \log {\cal L} + 0.3 \log Y.
\end{equation}
We define the luminosity ${\cal L} \equiv L/L_\odot$, mass ${\cal M}
\equiv m/\msun$ and radius ${\cal R} \equiv r/\rsun$ of the core of
the Wolf-Rayet star in solar units. $Y$ is the Helium mass fraction of
the star.

The effective escape velocity from the stellar core is
\begin{equation}
	\vesc \simeq 438 \sqrt{ {{\cal M} (1-\Gamma_{\rm e}) \over
	                         {\cal R}}} \; \; [\kms]
\end{equation}
Here $\Gamma_{\rm e} \simeq 7.66 \times 10^{-5} \sigma_{\rm e}{\cal
L}/{\cal M}$, corrects the gravity for electron scattering, and the
electron scattering coefficient is $\sigma_{\rm e} \simeq 0.4 (X +
\frac{1}{2}Y + \frac{1}{4}Z)$\,cm$^2$ (see Nugis \& Lamers, 2000).

The core radius is given by
\begin{equation}
	\log {\cal R} \simeq -1.845 + 0.338 \log {\cal L}.
\end{equation}
The mass is obtained by iterating the mass-luminosity relation for
massive Wolf-Rayet stars (Schaerer \& Maeder
1992)\nocite{1992A&A...263..129S}
\begin{equation}
	\log {\cal L} \simeq 2.782 + 2.695 \log {\cal M} - 0.461 
				\left( \log {\cal M} \right)^2.
\end{equation}

The $L-\vesc$ relation derived with this model fits well with WN4 to
WN6 stars of Hamann \& Koesterke (2000),\nocite{2000A&A...360..647H}
the cores of planetary nebul\ae\ of Kudritzki \& Puls
(2000)\nocite{2000ARA&A..38..613K} and with the sample of 24 Galactic
and 14 Magellanic Cloud O stars of Puls et al.\,
(1996)\nocite{1996A&A...305..171P}.
%
%

\begin{figure}[htbp!]
\hspace*{1cm}
\psfig{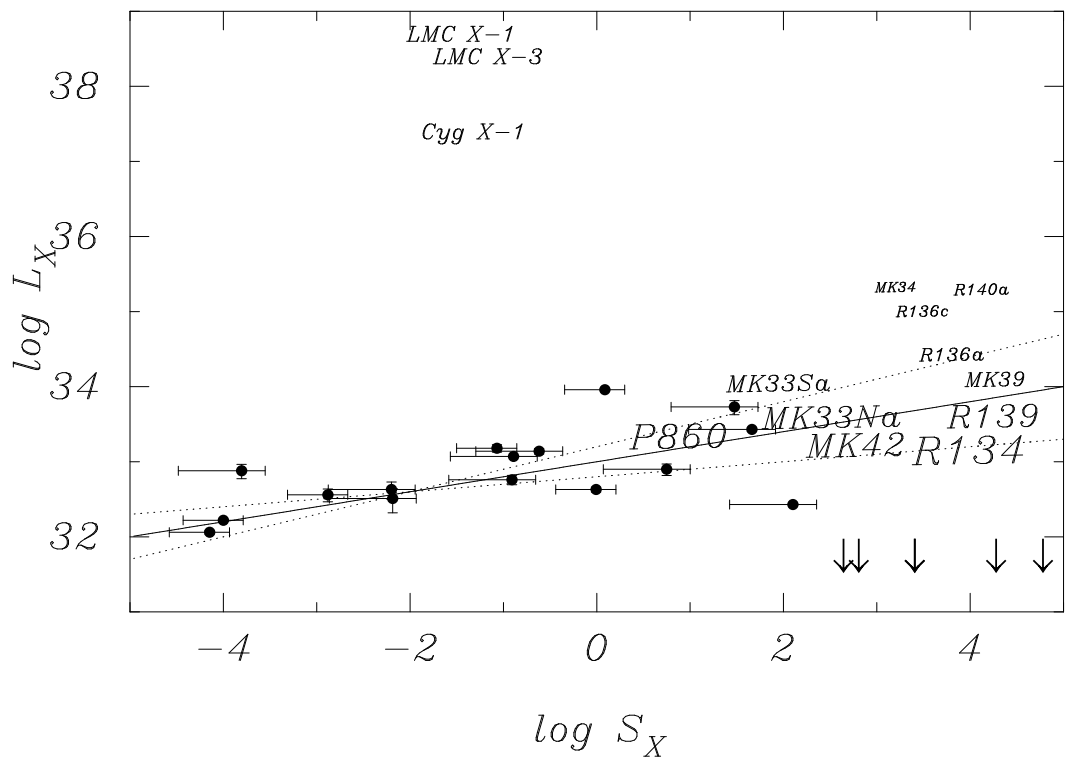}
\caption[]{X-ray luminosity as a function of $S_{\rm X}$
(Eq.\,\ref{Eq:Lxexp}). The bullets give the location of the binaries
listed by Chlebowski \& Garmany (1991).  The {\em Chandra} X-ray
sources from Tab.\,\ref{Tab:OC} are identified by their optical
counterpart; the size of their names are proportional to the $1\sigma$
errors.  The solid line gives a least squares fit to the bullets,
which represent the sources of Tab.\,\ref{Tab:KnownWR}.  The dotted
lines represent the uncertainty interval of the fit.  For comparison we
plot the location of three known high-mass X-ray binaries with a black
hole (see discussion).  The value of $S_{\rm X}$ (Eq.\,\ref{Eq:Lxexp})
for the five down pointed arrows represent the stars (from left to
right) R\,136b, Mk37Wb, Mk35, Mk30 and Mk49, which we did not detect 
in X-rays.  }
\label{Fig:fLx_WR}
\end{figure}

Fig.\,\ref{Fig:fLx_WR} shows the results of Eq.\,\ref{Eq:Lxexp} with
$S_{\rm X}$ along the horizontal axis and the observed X-ray
luminosity in the Einstein band along the vertical axis. The bullets
indicate the locations of the 16 colliding wind Wolf-Rayet binaries
from Chlebowski \& Garmany (1991).  The orbital separation for these
binaries is calculated using Keplers' third law with the mass and
orbital period given by Chlebowski \& Garmany (1991). When no mass
estimate is provided we assume a total binary mass of
$30\pm15$\,\msun.  A fit to this data is presented as the solid line
in Fig.\,\ref{Fig:fLx_WR}, which represents
\begin{equation}
 	L_{\rm X} = 10^{33.0\pm0.2} S_{\rm X}^{0.2\pm0.1},
\label{Eq:Lxfit}\end{equation}
where the dotted lines gives the uncertainty interval.

The names of the optical counterparts of the X-ray sources from
Tab.\,\ref{Tab:OC} are plotted in Fig.\,\ref{Fig:fLx_WR}.  The value
of $S_{\rm X}$ for these stars is calculated with Eq.\,\ref{Eq:Lxexp}
assuming that they are binaries with an orbital separation of
100\,\rsun. For the wind velocity in Eq.\,\ref{Eq:Lxexp}, we adopt
both extremes, the terminal velocity and the escape velocity from the
Wolf-Rayet star. The uncertainty thus obtained is proportional to the
size of the name tags in Fig.\,\ref{Fig:fLx_WR}.  Only the orbital
period of R\,140a is known (2.76\,days; Moffat et al. 1987).  With a
total binary mass of about 100\,\msun, this period corresponds to an
orbital separation of about 40\,\rsun. The accurate orbital separation
is not crucial for our crude model as it enters only linearly in
Eq.\,\ref{Eq:Lxexp}. However, when the orbital separation becomes
comparable to the size of the primary (windy) star, the wind velocity
\vwind\ may strongly deviate from the terminal wind velocity (via
Eq.\,\ref{Eq:vwind}). (Our assumption that binary members are
equipetomaniac and equally massive, has obviously limited validity.)

\begin{table*}[htbp!]
\caption[]{ Parameters for known black hole X-ray binaries and X-ray
sources with early type stars as optical counterparts. The first three
entries are the black hole X-ray binaries (Gies \& Bolton
(1982)\nocite{1982ApJ...260..240G}, and Lui et al.\,
2000)\nocite{2000A&AS..147...25L}.  The following 16 sources are X-ray
luminous O and B stars observed with {\em Einstein}. The X-ray data is
taken from Chlebowski, Harnden \& Sciortino
(1989)\nocite{1989ApJ...341..427C}. Temperature, luminosity, mass loss
rate, and wind velocities are from Chlebowski \& Garmany (1991).  }
\begin{flushleft}
\begin{tabular}{lc|crllclr}
\\ \hline
name & Spectral&$P_{\rm orb}$&\Teff&$\log L_{\rm bol}$&$\log L_X$ &$N_{\rm counts}$
       &$\log \Mwind$& $v_\infty$ \\
       & type  &  [days]
	&[1000K]&\multicolumn{2}{c}{[erg\,s$^{-1}$]}&
        &[\msunyr]& [km/s]\\
\hline 
Cyg X-1   & O9.7ab & 5.6   &     &     & 37.3    & var. & -7.0  & 1800 \\
LMC X-1   & O7--9II& 4.2   &     &     & 38.6    & var. & -7.0  & 2100 \\
LMC X-3   & B3V    & 2.3   &     &     & 38.6    & var. & -8.0  &  490 \\ 
\hline
HD 1337   & O9III  & 3.52  & 34.0&39.55& 32.90   & 32.8 &-6.17  & 2200 \\   
HD 12323  & ON9    &3.07   & 35.9&38.25& 32.88   & 21.1 &-8.37  & 1300 \\   
HD 37041  & O9V    &20.79  & 35.9&38.46& 32.06   & 305  &-8.24  & 1700 \\  
HD 37468  & O8.5III&30     & 35.0&38.82& 32.22   & 681  &-8.14  & 1300 \\     
HD 57060 & O7Ia   &4.39   & 36.1&39.81& 32.43   & 534  &-5.35  & 1800 \\   
HD 75759  & O9Vn   &33.31  & 35.9&39.23& 32.56   & 26.8 &-7.49  & 1600 \\  
HD 93206  & O9.7Ib &6      & 30.6&39.57& 33.43   & 327  &-5.85  & 2500 \\      
HD 93205  & O3V    &6.08   & 48.5&39.51& 33.14   & 144  &-6.23  & 3600 \\  
HD 93403  & O5III  &15.09  & 42.3&39.79& 33.96   & 667  &-5.82  & 3000 \\ 
HD 100213 & O8.5   &1.39   & 37.0&38.58& 32.63   & 15.1 &-7.65  & 1800 \\  
HD 152218 & O9.5IV &5.40   & 34.0&38.87& 32.76   & 48.9 &-6.93  & 3000 \\  
HD 152590 & O7.5V  &4.49   & 39.1&38.71& 32.51   & 7.95 &-7.36  & 2300 \\  
HD 159176 & O7V    &3.37   & 40.1&38.96& 33.07   & 859  &-6.72  & 2600 \\  
HD 165052 & O6.5V  &6.14   & 41.2&39.18& 33.18   & 77.5 &-6.62  & 2700 \\  
HD 206267 & O6.5V  &3.71   & 41.2&39.10& 32.63   & 120  &-6.16  & 3100 \\  
HD 215835 & O6V    &2.11   & 42.2&39.89& 33.73   & 22.1 & -5.54  & 3350 \\ 
\hline
\end{tabular}
\end{flushleft}
\label{Tab:KnownWR} 
\end{table*}

\section{Discussion}\label{Sect:Disc}

We have analyzed archival {\em Chandra} data of the central portion of
the 30\,Doradus region including the star clusters R\,136 and
R\,140.

We confirm the detection of two bright X-ray sources with the stars
Mk34 and R\,140a as optical counterparts (see Wang \& Helfand
1991).\nocite{1991ApJ...370..541W} The X-ray luminosities of these two
sources have not noticeably changed in the 9 years between
observations.  The difference in X-ray luminosity between
the earlier observation and ours can be attributed to the different
spectral fitting.

We identified 20 point sources with an X-ray luminosity ranging from
$5\times 10^{32}$ to $2\times 10^{35}$\,erg\,s$^{-1}$.  Nine point
sources are located within 14$''$ (3.4\,pc) of the center of the star
cluster R\,136, one possibly extended source is associated with the
cluster of WN stars R\,140. Four others have WN stars as counterparts
which do not seem to be associated with either of the two
clusters. 

Wang (1995) analyzed ROSAT spectra of our sources CX5 (his source
number 7) and CX10 (his number 5).  He concluded that they are
probably high-mass X-ray binaries with a black hole as accreting star.
His arguments are based on the spectral temperature (0.1--0.2\,keV for
his source number 5 and 0.2--0.4\,keV for number 7) and the high
luminosity ($\sim 10^{36}$\,erg\,s$^{-1}$).  In addition he uses the
measured orbital period of 2.76 days and mass function of 0.12\,\msun\
of the star R\,140a2 (see Moffat et al. 1987) as an argument for the
presence of a dark object.


For comparison we calculate the value of $S_{\rm X}$
(Eq.\,\ref{Eq:Lxexp}) for the three known black hole X-ray binaries
(see Tab.\,\ref{Tab:KnownWR}) and plot these in
Fig.\,\ref{Fig:fLx_WR}. Their positions in Fig.\,\ref{Fig:fLx_WR} are
completely different from the X-ray sources in NGC\,2070.  Of course,
there is an interpretational difficulty here about which state of the
black hole binary (see Tanaka \& Lewin 1995)\nocite{1995Tanaka&Lewin}
should be used for a comparison.  In any case, the age of the cluster
would make it remarkable if any of the X-ray sources were associated
with a black hole or a neutron star. On these grounds we conclude that
none of the X-ray sources in NGC\,2070 are associated with compact
objects.

We conclude that all sources are consistent with colliding wind
systems.  X-rays are produced in the collisions between the winds in
close binaries in which these stars reside. We derive an empirical
relation for the X-ray luminosity for a colliding wind binary by
substitution of Eq.\,\ref{Eq:Lxexp} in Eq.\,\ref{Eq:Lxfit}:
\begin{equation}
      L_{\rm x} = 1.3 \times 10^{34}  
  	        \left( {\Mwind \over 10^{-5}\msun {\rm yr}^{-1}} \right)^{0.4}
		\left( {\rm 1000 km s}^{-1} \over \vwind \right)^{0.65} 
		\left( {\rsun \over a} \right)^{0.20} \;\; [{\rm erg\, s}^{-1}]
\label{Eq:Lx}\end{equation}
This relation has a weaker dependence on the velocity and orbital
separation compared to Usov's (1992) theoretical expression (his
Eq. 89).

About $70$\,\% (13/18) of the Wolf-Rayet stars in NGC\,2070 are bright
in X-rays.  The star cluster also contains 5 spectral type O3f$^\star$
and WN stars (R\,136b, Mk37Wb, Mk35, Mk30 and Mk49) which do not show
up in our X-ray image.  Using Eq.\,\ref{Eq:Lx} one would expect X-ray
fluxes at least an order of magnitude above our detection threshold of
$5\times 10^{32}$\ erg\,s$^{-1}$ (see down pointed arrows in
Fig.\,\ref{Fig:fLx_WR}).  The absence of X-rays for these stars can be
explained by the companion mass being much smaller than that of the WN
star or by the orbital separation being much larger than the adopted
100\,\rsun.  Alternatively, these stars are single, in which case they
may be much dimmer in X-rays.

The star cluster NGC\,3603 has characteristics similar to R\,136, and
therefore may also host a wealth of X-ray sources.  As suggested by
the referee, we examined the public 50\,ks {\em Chandra} data centered
on NGC\,3603, but it goes beyond the scope of this paper to fully
reduce and analyze these data.  The cluster shows a wealth of
interesting features including diffuse X-ray emission, exceptionally
bright point sources and a large number of much dimmer point sources
(Moffat, Corcoran, Stevens, Skalkowski, Marchenko, M\"uke, Ptak,
Koribalski, Mushotzky, Pittard, Pollock and Brandner, 2002 in
preparation and Stevens, private communication).  The sources in the
cluster center are heavily blend and it will be a very time-consuming
task to find optical counterparts.


Other clusters with similar characteristics as R\,136 are the Arches,
the Quintuplet cluster and Westerlund 1.  These clusters also contain
a wealth of bright and young early spectral types star which, when in
binaries, may be bright X-ray point sources with characteristics
similar to those discussed in relation to R\,136.

\section{Conclusions}\label{Sect:Concl}

We analyzed a 21\,ksec archival {\em Chandra} X-ray observation
pointed at the central portion of the 30 Doradus region in the Large
Magellanic Cloud.  The image contains 18 new X-ray sources and we
confirm the existence of two sources earlier discovered by Wang \&
Helfand (1991).\nocite{1991ApJ...370..541W} Nine sources are within
$\sim 3.4$\,pc of the center of the young star cluster R\,136.  The
X-ray luminosity (0.2 -- 3.5\ keV) of these sources ranges from
$5\times 10^{32}$ (our detection threshold) to $2 \times 10^{35}$
erg\,s$^{-1}$.  The two known sources have not changed noticeably in
X-ray luminosity over the 9 years between the {\em Einstein} and our
{\em Chandra} observations.

We conclude that all observed sources, except for 5 unidentified
sources, the IRAS source IRS\,2, and the foreground infrared source
MH\,2, have stars as counterparts (10 WN Wolf-Rayet, two Of$^\star$
and one O3V star).  We argue that the X-rays in these stars are
produced by colliding winds.

We do not agree with Wang (1995) that the two earlier discovered
sources (CX5 and CX10) host black holes because 1) the stellar
environment is too young to produce black holes, 2) the spectra and
X-ray luminosities of the sources do not at all agree with other known
black hole candidates, and 3) the sources fit well with our simple
semi-analytic colliding wind model.

The X-ray luminosities of the other observed sources agree well with
our simple colliding wind model.  The model gives the X-ray luminosity
as a function of the stellar wind mass loss, its terminal velocity and
the binary separation.  The empirical fit, calibrated to 16 known
colliding wind binaries, culminates in Eq.\,\ref{Eq:Lx}.

This empirical relation can be used to reduce the noise in the
relation between the 0.2 to 3.5\ keV X-ray luminosity and the
Bolometric luminosity of early type stars. The luminosity calculated
with Eq.\,\ref{Eq:Lx} provides a powerful diagnostic for studying
colliding wind X-ray binaries.

If the X-rays are indeed the result of colliding winds in close
binaries, possibly all 13 identified sources detected with {\em
Chandra} could be close binaries.  In that case we conclude that the
binary fraction among early type stars in the 30\,Doradus region is
unusually high; possibly all early type stars are binaries.

Other star clusters, with age, mass and concentration similar to
R\,136 are likely to contain a similar wealth in bright X-ray
sources. In these cases, the emission may also be produced by
colliding stellar winds.  We then conclude that the star cluster
NGC\,3603 may contain more than 20 X-ray sources brighter than
$10^{33}$ erg\,s$^{-1}$ but the majority may be blended in the cluster
center.  (Preliminary results of an X-ray study with {\em Chandra}
ware reported at the AAS December meeting by Corcoran et al., 2000,
Moffat et al.\, 2002 in preparation)\nocite{AASnonsence}

\acknowledgements We are grateful to Ron Remillard and Ian Stevens for
discussions. This work was supported by NASA through Hubble Fellowship
grant HF-01112.01-98A awarded (to SPZ) by the Space Telescope Science
Institute and by the Royal Dutch Academy or Sciences (KNAW).  DP
acknowledges that this material is based upon work partially supported
under a National Science Foundation Graduate Fellowship.

\end{document}